\documentclass[paper,notoc,nohyper]{JHEP}
\usepackage{epsfig}
\usepackage{amsbsy}

\title{\bf Progress in NNLO calculations for scattering processes}

\author{E.~W.~N.~Glover\footnote{Based on invited talks at
6th International Symposium on Radiative Corrections,
Application of Quantum Field Theory to Phenomenology (RADCOR 2002)
and 6th Zeuthen Workshop on Elementary Particle Theory
(Loops and Legs in Quantum Field Theory)
Kloster Banz, Germany,
8th - 13th Sep 2002
and
14th Topical Conference on Hadron Collider Physics (HCP 2002), Karlsruhe, Germany,
29th Sep - 4th Oct 2002.
}
\\
{Institute for Particle Physics Phenomenology, University of Durham \\
        South Road, Durham, DH1 3LE, U.K.\\[1mm] E-mail: \email{E.W.N.Glover@durham.ac.uk}}%
        }

\abstract{
The various motivations for improving the perturbative prediction to
next-to-next-to-leading order (NNLO) for basic
scattering processes in proton-(anti)proton, electron-proton and
electron-positron scattering are discussed in detail.  Recent progress in the
field of next-to-next-to-leading order calculations is reviewed.
}

\preprint{{DCTP/02/146}, {IPPP/02/73}, {hep-ph/0211412}}

\begin{document}

\section{Why NNLO calculations are important.}

In the past decade, QCD has become a quantitative science and comparisons  of
experimental data with NLO QCD have become the standard. However, there are many
reasons why extending perturbative calculations to NNLO is vital in reducing
the theoretical uncertainty.  In the following we list six of them.

\subsection{Renormalisation scale uncertainty}

In many cases, the uncertainty from the pdf's and from the choice of the
renormalisation scale $\mu_R$ give uncertainties that are as big as or bigger than
the experimental errors.  Of course, the theoretical prediction should be
independent of $\mu_R$.   However, a scale dependence is introduced by
truncating the perturbative series. The change due to varying the scale is
formally higher order.
If an observable ${\cal O}bs$ is known to order $\alpha_s^N$,
\[
{\cal O}bs=\sum_{0}^N A_n(\mu_R) \alpha_s^n(\mu_R),\nonumber
\]
then,
\[
\frac{\partial}{\partial \ln(\mu_R^2)} 
{\cal O}bs = {\cal O} \left(\alpha_s^{N+1} \right).\nonumber
\]
Often the uncertainty due to uncalculated higher
orders is estimated by varying the renormalisation scale upwards and downwards
by a factor of two around a typical hard scale in the process.  
However, the variation only produces copies of the lower order terms,
\\
e.g. \begin{eqnarray*}
{\cal O}bs &=& A_1 \alpha_s(\mu_R) \\
&+& \left(A_2 + b_0 A_1
\ln\left({\mu_R^2\over \mu_0^2}\right)\right)
\alpha_s^2(\mu_R).
\end{eqnarray*}
$A_2$ will generally contain infrared logarithms and constants that are
not present in $A_1$ and  therefore {\em cannot be predicted} by varying
$\mu_R$. For example, $A_1$ may contain infrared logarithms $L$ up to $L^2$,
while $A_2$ would contain these logarithms up to $L^4$.  $\mu_R$ variation is
{\em only an estimate} of higher order terms and a large variation probably means
that {\em predictable} higher order terms are large.

To illustrate the improvement in scale uncertainty that may occur at NNLO, let
us consider the production of a central jet in $p\bar p$ collisions.   The
renormalisation scale dependence is entirely predictable,
\begin{eqnarray*}
\lefteqn{\frac{d\sigma}{dE_T} = \alpha^2_s(\mu_R) {A_0}}\\
&+&  \alpha^3_s(\mu_R) \left ({A_1} + 2 b_0 L {A_0}\right)\\
&+&  \alpha^4_s(\mu_R) \left ({A_2} + 3 b_0 L {A_1} + (3b_0^2L^2+2 b_1 L) { A_0} \right)
\end{eqnarray*}
with $ L = \log(\mu_R/E_T)$.  $A_2$ and $A_3$ are the known LO and NLO coefficients
while $A_4$ is the presently unknown NNLO term.  Inspection of Fig.~1 shows
that the scale dependence is systematically reduced by increasing the number of
terms in the perturbative expansion.   At NLO, there is always a turning point
where the prediction is insensitive to small changes in $\mu_R$.  If this
occurs at a scale far from the typically chosen values of $\mu_R$, the
$K$-factor (defined as $ K = 1 + \alpha_s(\mu_R) {A_3}/{A_2}$) will be
large.  At NNLO the scale dependence is clearly significantly reduced, although
a more quantitative statement requires knowledge of $A_4$. 

\begin{figure}[t]
\label{fig:nnlo}
\begin{center}
\includegraphics[height=8cm]{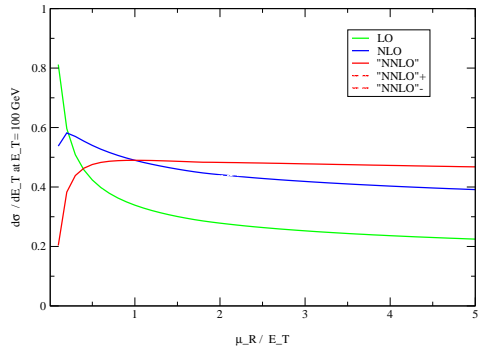}
\end{center}
\vspace{-4cm}
\caption{Single jet inclusive distribution at $E_T = 100$~GeV and $0.1 < |\eta| < 0.7$ at
$\sqrt{s} = 1800$~GeV at LO (green), NLO (blue) and NNLO (red). The solid and
dashed red lines
how the NNLO prediction if $A_4=0$, $A_4= \pm A_3^2/A_2$ respectively. 
The same pdf's and $\alpha_s$ are used throughout.}
\end{figure}

\subsection{Factorisation scale dependence}

Similar qualitative arguments can be applied to the factorisation scale inherent
in perturbative predictions for quantities with initial state hadrons. 
Including the NNLO contribution reduces the uncertainty due to the truncation of
the perturbative series.

\subsection{Jet algorithms}
There is also a mismatch between the number of hadrons and the number of
partons in the event.   At LO each parton has to model a jet and there is no
sensitivity to the size of the jet cone.   At NLO two partons can combine
to  make a jet giving sensitivity to the shape and size of the jet cone.
Perturbation theory is starting to reconstruct the parton shower within the
jet. This is further improved at NNLO where up to three partons can form a
single jet, or alternatively two of the jets may be formed by two partons. This
should lead to a better matching of the jet algorithm between theory and 
experiment.

\subsection{Transverse momentum of the incoming partons}
At LO, the incoming particles have no transverse momentum with respect to the
beam so that the final state is produced at rest in the transverse plane.  At
NLO, single hard radiation off one of the incoming particles gives the final
state a transverse momentum kick even if no additional jet is observed. In some
cases, this is insufficient to describe the transverse momentum distribution
observed in the data and one appeals to the
intrinsic transverse motion of the partons confined in the proton to provide an
enhancement.    However, at NNLO, double radiation from one particle or  single
radiation off each incoming particle gives more complicated transverse momentum
to the final state and should provide a better, and more theoretically motivated, 
description of the data.

\subsection{Power corrections}
Current comparisons of NLO predictions with experimental data generally reveal
the need for power corrections.   For example, in
electron-positron annihilation,  the experimentally measured average value of
1-Thrust lies well above the NLO predictions.   The difference can be accounted
for by a $1/Q$ power correction. While the form of the power correction can be
theoretically motivated, the magnitude is generally extracted from data and, to
some extent, can be attributed to uncalculated higher orders.   Including the
NNLO may therefore reduce the size of the phenomenological  power correction
needed to fit the data.

\begin{figure}[t]
\label{fig:powerc}
\begin{center}
\includegraphics[width=5.8cm]{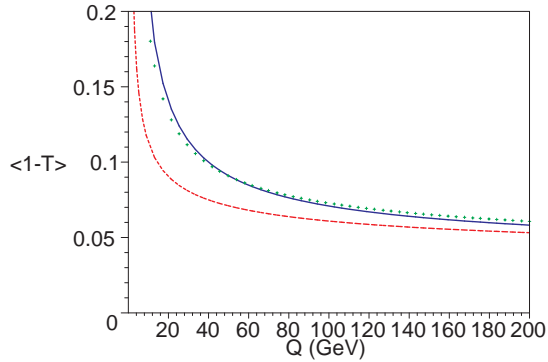}
\end{center}
\caption{The average value of $\langle 1-T\rangle$ given by Eq.~\ref{eq:omt}
showing the NLO prediction (dashed red), the NLO prediction with power correction of
$\lambda = 1$~GeV (solid blue) and an NNLO estimate with $A_3=3$ and a power correction
of $\lambda=0.5$~GeV (green dots).}
\end{figure}
Before the calculation of the NNLO contribution it is not possible to make a more
quantitative statement. However to illustrate the qualitative point, 
let us take the simple example of an observable like
$\langle 1-T\rangle$ which can be modelled by the simplified series,
\begin{equation}
\label{eq:omt}
\langle 1-T\rangle = 0.33 \alpha_s(Q) + 1.0 \alpha_s(Q)^2 + A_3 \alpha_s(Q)^3 +
\frac{\lambda}{Q},
\end{equation}
with $\alpha_s(Q) \sim 6\pi/23/\log(Q/\Lambda)$ and $\Lambda = 200$~MeV. 
Fig~2 shows the NLO
perturbative prediction $A_3=0$,  $\lambda=0$ as well as the NLO prediction
combined with a power correction, $A_3=0$, $\lambda=1$~GeV which can be taken to
model the data.   If the NNLO coefficient turns out to be positive (which is by
no means guaranteed), then the size of the power correction needed to describe
the data will be reduced. For example, if we estimate the NNLO coefficient as
$A_3 = 3$, which is large but perhaps not unreasonable, then the NLO prediction
plus power correction can almost exactly be reproduced with a power correction
of the same form, but $\lambda  = 0.6$~GeV.  We are effectively trading a
contribution of ${\cal O}(1/Q)$ for a contribution of $1/\log^3(Q/\Lambda)$.  
At present the data is insufficient to distinguish between these two functional
forms. 

\subsection{The shape of the prediction}
It is an oft repeated statement that NLO predictions affect the normalisation
of the leading order prediction but not the shape of the distribution.   This
is wrong for two reasons.  First, as indicated in Table.~1, the loop
contribution gets new kinematic contributions.  Of course, each ratio of
kinematic scales is further multiplied by logarithms and polylogarithms of
ratios of scales.  Second, as the number of final state particles increases,
the allowed phase space is enlarged - leading, for example, to lower values of thrust in
electron-positron annihilation and jet production at higher rapidities in
hadron-hadron collisions.   These effects simply cannot be modelled by scaling
the lowest order cross section.  
\begin{table}[t]
\begin{center}
\begin{tabular}{ cccc }
Kinematic scales  &   Tree & One-loop & Two-loop
\\
\hline
&&&\\
$\frac{(u^2+t^2)}{s^2}$ & $\surd$ & $\surd$ & $\surd$ \\
&&&\\
$\frac{(u^2-t^2)}{s^2}$, 1 &   & $\surd$ & $\surd$ \\
&&&\\
$\frac{u^3}{s^2t}$, $\frac{t^3}{s^2u}$ & & &$\surd$ \\
&&&\\
\hline
\end{tabular}
\end{center}
\caption{Ratios of kinematic scales for the scattering of unlike massless quarks, $q \bar
q \to q^\prime \bar q^\prime$}
\end{table}

\subsection{Parton densities at NNLO}
\label{sec:ho;nnlopdf}

Consistent NNLO predictions for processes involving hadrons in the  initial
state require not only the NNLO hard scattering  cross sections, but also
parton distribution functions which are accurate to this order. 

The evolution of parton distributions is governed at NNLO by the three-loop
splitting functions, which are not fully known at present, although the
expectation is that they will be determined shortly, see Ref.~\cite{Vermaseren:2002rn}
and references theirin.  

The determination of NNLO parton distributions requires a global  fit to the
available data on a number of hard scattering  observables, with all
observables computed consistently at NNLO.  At present, the NNLO coefficient
functions are available for  the inclusive  Drell-Yan
process~\cite{Hamberg:1990np,Harlander:2002wh}  and for deep inelastic
structure  functions~\cite{Zijlstra:1992qd}. These two observables are by
themselves  insufficient to fully constrain all parton species. In particular
large transverse energy jet production in hadron-hadron collisons provides a
vital constraint on the gluon at large $x$ and therefore requires the
determination of the single jet inclusive cross section at NNLO.

\section{Recent progress in the field}

There has been enormous progress in calculating two-loop matrix elements since
the previous ``Loops and Legs" workshop in April 2000.  To a large extent this
has been due to breakthoughs in the calculation of two-loop master integrals
with massless propagators.  In tour de force calculations through the summer of
1999, Smirnov~\cite{Smirnov:1999gc} and Tausk~\cite{Tausk:1999vh} computed the
planar and non-planar  double box master integrals with four on-shell legs
using Mellin Barnes integrals.  The other four  four-point master integrals
needed for the on-shell case were much easier to calculate   
~\cite{Smirnov:1999wz,Anastasiou:1999bn,Anastasiou:2000mf,Gehrmann:2000xj,Anastasiou:2000kp}.

Smirnov has applied the same Mellin Barnes technique to the case where one leg
is off-shell~\cite{Smirnov:2000vy,Smirnov:2000ie}, however, the main break
through here is due to Gehrmann and Remiddi~\cite{Gehrmann:2000zt,Gehrmann:2001ck}
who have provided expansions for all of the relevant master integrals by making use of
differential equations for two-loop four point
functions~\cite{Gehrmann:1999as}.   For integrals with singularities through to
$1/\epsilon^4$ it proved necessary to introduce two-dimensional harmonic
polylogarithms in addition to the standard Nielsen polylogarithms.   Numerical
evaluations of the master integrals~\cite{Gehrmann:2001pz,Gehrmann:2001jv} and
analytic continuations have been provided in Ref.~\cite{Gehrmann:2002zr}.

The next major step is to reduce the tensor integrals appearing in Feynman
diagram calculations to the few master integrals for which series expansions
are known.   The main tool here is integration-by-parts 
(IBP)~\cite{Chetyrkin:pr,Chetyrkin:qh} which, for two-loop four point
integrals, generates 10 equations relating integrals with different powers of
propagators.  To this can be added the three Lorentz Invariance (LI)
identities~\cite{Gehrmann:1999as}.   Together, these equations represent the
fact that loop integrals exhibit Poincare invariance. Together, the IBP and LI
equations form a linear system that can be automatically solved with algebraic
methods to relate all tensor integrals to the master
integrals~\cite{Gehrmann:1999as,Laporta:2001dd}.  

These technological tools have allowed the computation of  a whole raft of
two-loop matrix elements for scattering processes with up to one off-shell leg
as detailed in Table~2.
\begin{table}[t]
\begin{center}
\begin{tabular}{c c c}
On-shell Process  &   Tree    &  Helicity  \\
   &     $\times$ Two-loop &    amplitudes \\
\hline
${e^+e^- \to \mu^+\mu^- (e^+e^-)}$ & \cite{Bern:2000ie} &   \\
${ q\bar q \to q\bar q (\bar q^\prime q^\prime)}$ & \cite{Anastasiou:2000kg,Anastasiou:2000ue} &   \\
${ q \bar q \to gg}$ & \cite{Anastasiou:2001sv}&        \\
${ gg\to gg}$ & \cite{Glover:2001af} & \cite{Bern:2000dn,Bern:2002tk} \\
${ gg\to \gamma\gamma}$ & --- & \cite{Bern:2001df}  \\
${ \gamma\gamma\to \gamma\gamma}$ & ---  & \cite{Bern:2001dg,Binoth:2002xg} \\
${ q\bar q \to g\gamma (\gamma\gamma)}$ & \cite{Anastasiou:2002zn} &    \\
\hline &&\\

Off-shell Process  &   & \\\hline
${e^+e^- \to q\bar q g}$ & \cite{Garland:2001tf} &  \cite{Garland:2002ak,Moch:2002hm}
\\
\hline
\end{tabular}
\end{center}
\caption{Current status of available two-loop amplitudes for scattering
processes with all legs on-shell and with one leg off-shell}
 \end{table}

A vital check on the infrared structure of these two-loop matrix elements is
given by Catani~\cite{Catani:1998bh}.  In Catani's formalism, the singular
${\cal O}(1/\epsilon^4)$, ${\cal O}(1/\epsilon^3)$, ${\cal O}(1/\epsilon^2)$
structure can be straightforwardly be determined while the non-logarithmic
coefficient $H_2$ multiplying the ${\cal O}(1/\epsilon)$  can also be
extracted.   For some time, the origins of this formula were lost in transit, 
however Sterman and Tejeda-Yeomans~\cite{Sterman:2002qn} have now
rederived it from the basis of factorisation of inter-jet and intra-jet
radiation and given a more concrete explanation of how $H_2$ is produced.

\section{What remains to be done}

Of course knowledge of the two-loop amplitudes is only one of the ingredients 
needed in the construction of a  NNLO parton level Monte Carlo.     In addition
we also need; 
\begin{itemize} 
\item square of the one-loop $2 \to 2$ amplitudes, \vspace{-0.2cm}
\item interference of tree and one-loop $2 \to 3$ amplitudes,  \vspace{-0.2cm}
\item square of the tree-level $2 \to 4$ amplitudes.   
\end{itemize}
The tree level $2 \to 4$ amplitudes have been known for some time, while the 
five point amplitudes with zero~\cite{Bern:1993mq,Bern:1994fz,Kunszt:1994tq} and 
one~\cite{Glover:1996eh,Bern:1996ka,Campbell:1997tv,Bern:1997sc} off-shell legs have also been worked
out.

These latter two processes contribute when either one or two of the partons are
unresolved and there is a   
much more sophisticated infrared cancellation 
between 
\begin{itemize}
\item 
$n$ and $n+1$ particle
contributions when one particle is unresolved,\vspace{-0.2cm}
\item 
$n$ and $n+2$ particle
contributions when two particles are unresolved.
\end{itemize}
At present, a detailed method for subtracting the singularities and numerically
evaluating the finite remainders has not been worked out.

Nevertheless, the prognosis is favourable.   If the rate of technical
development maintains its current pace, it is extremely likely that by the time
of the next workshop, NNLO Monte Carlo's will be available for some of the most
basic scattering processes, leading to a more complete and quantitative
understanding of hard interactions.
 
\section*{Acknowledgements}
I am very happy to thank Babis Anastasiou, Thomas Binoth,  Thomas Gehrmann, Lee
Garland, Thanos Koukoutsakis, Peter Marquard, Carlo Oleari,  Ettore Remiddi,
Maria-Elena Tejeda-Yeomans and  Jochum van der Bij for many stimulating
collaborations in the past three years.

\end{document}